\documentclass[pdflatex,sn-mathphys-num]{sn-jnl}

\usepackage{graphicx}%
\usepackage{multirow}%
\usepackage{amsmath,amssymb,amsfonts}%
\usepackage{amsthm}%
\usepackage{mathrsfs}%
\usepackage[title]{appendix}%
\usepackage{xcolor}%
\usepackage{textcomp}%
\usepackage{manyfoot}%
\usepackage{booktabs}%
\usepackage{algorithm}%
\usepackage{algorithmicx}%
\usepackage{algpseudocode}%
\usepackage{listings}%

\theoremstyle{thmstyleone}%
\newtheorem{theorem}{Theorem}
\theoremstyle{thmstyletwo}%
\newtheorem{example}{Example}%
\newtheorem{remark}{Remark}%

\theoremstyle{thmstylethree}%
\newtheorem{definition}{Definition}%

\newcommand{\RR}{\mathbb{R}}
\newcommand{\R}{\mathbb{R}}

\newcommand{\Tn}{\mathcal{U}_m}

\newcommand{\Trop}{\text{Trop}}

\def\R{{\mathbb R}}

\begin{document}

\title[KDE Treespace]{Bandwidth Selection of Density Estimators over Treespaces}


\author*[1]{\fnm{Ruriko} \sur{Yoshida}}\email{ryoshida@nps.edu}
\author[2]{\fnm{Zhiwen}\sur{Wang}}\email{zw26@njit.edu}



\affil*[1]{\orgdiv{Department of Operations Research}, \orgname{Naval Postgraduate School}, \orgaddress{\street{1411 Cunningham Road}, \city{Monterey}, \postcode{93943}, \state{CA}, \country{USA}}}
\affil[2]{\orgdiv{Department of Mathematical Science}, \orgname{New Jersey Institute of Technology}, \orgaddress{\street{University Heights }, \city{Newark}, \postcode{07102}, \state{NJ}, \country{USA}}}



\abstract{{\color{black}
A kernel density estimator (KDE)
is one of the most popular non-parametric density estimators. In this paper we focus on a
best bandwidth selection method for use in an analogue of a classical KDE using the tropical
symmetric distance, known as a tropical KDE, for use over the space of phylogenetic trees. We
propose the likelihood cross validation (LCV) for selecting the bandwidth parameter for the KDE
over the space of phylogenetic trees.  In this paper, first, we show the explicit optimal solution of the best-fit bandwidth parameter via the LCV for tropical KDE over the space of phylogenetic trees.  
Then, computational experiments with simulated datasets generated under the multi-species coalescent (MSC) model show that a tropical KDE with the best-fit bandwidth parameter via the LCV perform better than a tropical KDE with an estimated best-fit bandwidth parameter via nearest neighbors in terms of accuracy and computational time.   Lastly, we apply our method an empirical data from the Apicomplexa genome.}}




\maketitle

\section{Introduction}\label{sec1}
{\color{black} Phylogenomics is a relatively new field in evolutionary biology. It applies tools in phylogenetics to genome data,with the goal of identifying genes whose evolutionary histories differ from the evolutionary history of the given set of species.} Such evolutionary processes include negative or balancing selection on a locus, which might increase the chance for ancestral gene copies to maintain through speciation events \cite{Takahata}, and horizontal gene transfer, which shuffles divergent genes among different species \cite{Liu2007}.  

To find evolutionary history of each gene in a genome data, we often infer a {\em gene tree} which is a {\em phylogenetic tree} reconstructed from an alignment of the gene.  A phylogenetic tree is a weighted (rooted or unrooted) tree representation of an evolutionary history of the given set of species whose leaves represent species and internal nodes are unlabeled.  Each branch in a phylogenetic tree has a non-negative weight.  In this paper we focus on rooted gene trees.  More specifically, we focus on an {\em equidistant tree}, a rooted phylogenetic tree whose total branch lengths from its root to each leaf is same.  

A {\color{black}k}ernel density estimator (KDE) is one of the most popular non-parametric density estimator.  In 2024, Yoshida et al.~in \cite{tropicalKDE} introduced a {\em tropical KDE} to estimate a distribution of gene trees over the {\em space of phylogenetic trees} with a fixed number of leaves $m$ {\color{black}using} the {\em tropical symmetric distance} $d_{\rm tr}$.  The space of phylogenetic trees {\color{black}with $m$ leaves} is the set of all possible phylogenetic trees with the leaf label set {\color{black}$[m]:= \{1, \ldots , m\}$. It is well-known} that the space of phylogenetic trees is the union of $m-2$ dimensional polyhedral cones over $\mathbb{R}^{\binom{m}{2}}$ (for example see \cite{BHV}){\color{black}, and thus, is not convex. } Therefore we cannot directly apply classical density estimators using traditional kernels, such as the {\color{black}Gaussian or Laplacian} kernel functions.   Weyenberg et al.~in \cite{KDE} used the Billera-Holmes-Vogtmann (BHV) metric developed by Billera, Holmes, and Vogtmann in \cite{BHV} to define a kernel density estimator over the space.  However they also showed that the normalized constant for the kernel function with the BHV metric over the space of phylogenetic tree is not constant, i.e., it varies {\color{black}depending} on the location of the function.  In 2016, Weynberg et al.~in \cite{KDETree} {\color{black}developed} an algorithm to approximate the normalized constant of the kernel function with the BHV metric using the holonomic gradient methods.  {\color{black}However,the resulting normalized constant approximation came at significant computational cost. } In 2024, Yoshida et al.~used the fact that the space of all equidistant trees is {\em tropically convex} shown by Ardila and Develin in \cite{Ardila} and used the tropical symmetric distance $d_{\rm tr}$ to define a kernel function.  Yoshida et al.~in \cite{tropicalKDE} showed that the normalized constant of their kernel function is constant by computational experiments. {\color{black}To} compute a bandwidth of each kernel function, they used simply a nearest neighbor to estimate the best-fit bandwidth. R code is available at \url{https://polytopes.net/tropical_KDE_new.zip}. 

In this paper we first modify a tropical KDE proposed by Yoshida et al.~in \cite{tropicalKDE}.  Then, we propose an analogue of the likelihood cross validation (LCV) defined by Habbema et al.~\cite{habbema1974stepwise} in terms of $d_{\rm tr}$ and we show the {\color{black}explicit } global optima in terms of the LCV with $d_{\rm tr}$.   For computational experiments, {\color{black}first,}
we apply {\color{black}tropical} KDEs {\color{black} with the best-fit bandwidth parameter via the LCV} to simulated datasets generated by the multi-species coalescent (MSC) model {\color{black} and our computational experiments show that our KDEs with the best-fit bandwidth parameters via the LCV perform better than KDEs proposed by Yoshida et al.~in \cite{tropicalKDE} in terms of accuracy and computational time.  Lastly, we apply our method to}  Apicomplexa genome dataset generated by Kuo et al.~\cite{kuo}. 

\section{Basics}

Let $e:= \binom{m}{2}$.  Throughout this paper, we consider the {\it tropical projective torus} $\mathbb R^e \!/\mathbb R {\bf 1}$, {which} is isomorphic to $\R^{e-1}$.
For more details, see \cite{ETC,MS}.

\subsection{Tropical Basics}

\begin{definition}[Tropical Arithmetic Operations]
Under the tropical semiring $(\,\mathbb{R} \cup \{-\infty\},\boxplus,\odot)\,$, we consider the {\em max-plus arithmetic}, that is, the tropical arithmetic operations of addition and multiplication defined as:
$$a \boxplus b := \max\{a, b\}, ~~~~ a \odot b := a + b,$$
where $ a, \, b \in \mathbb{R}\cup\{-\infty\}.$
Notice that over the tropical semiring with the max-plus arithmetic, the identity element under the tropical addition $\boxplus$ is $-\infty$ and the identity element under tropical multiplication $\odot$ is $0$.
\end{definition}

\begin{definition}[Tropical Scalar Multiplication and Vector Addition]
For any scalars $a,\, b \in \mathbb{R}\cup \{-\infty\}$ and for any vectors $v = (v_1, \ldots ,v_e), w= (w_1, \ldots , w_e)$ over the {\em tropical projective space} $\left((\mathbb{R}\cup-\{\infty\})^e - \{(-\infty, \ldots , -\infty)\}\right)\!/\mathbb R {\bf 1}$, we have tropical scalar multiplication and tropical vector addition as:
$$a \odot v \boxplus b \odot w := (\max\{a+v_1,b+w_1\}, \ldots, \max\{a+v_e,b+w_e\}).$$
\end{definition}
Note that the tropical projective space $\left((\mathbb{R}\cup-\{\infty\})^e - \{(-\infty, \ldots , -\infty)\}\right)\!/\mathbb R {\bf 1}$ contains the tropical projective torus $\mathbb R^e \!/\mathbb R {\bf 1}$ and {\color{black}throughout }this paper we consider the following equivalence relations: for any vector $v = (v_1, \ldots ,v_e)$ over $\mathbb R^e \!/\mathbb R {\bf 1}$ and for a fixed {\color{black}height} $h >0$, we have
\[
(v_1, \ldots ,v_e) = (v_1 - h_0, \ldots ,v_e-h_0)
\]
where $h_0 = \max_{i \in [e]} v_i - h$ so that we {\color{black} always have} $\max_{i \in [e]} v_i = h$. 

\begin{definition}\label{def:polytope}
Suppose we have a set $S \subset \mathbb R^e \!/\mathbb R {\bf 1}$. 
$S$ is {\em tropically convex} if
\[
c_1 \odot v \boxplus c_2 \odot w \in S
\]
for any $c_1, c_2 \in \R$ and for any points $v, w \in S$. 
Suppose $V = \{v^1, \ldots , v^s\}\subset \mathbb R^e \!/\mathbb R {\bf   1}$.  The smallest tropically-convex subset containing $V$ is called the {\em tropical convex hull} or {\em tropical polytope} of $V$ which can be written as the set of all tropical linear combinations of $V$ as:
$$ \mathrm{tconv}(V) = \{a_1 \odot v^1 \oplus a_2 \odot v^2 \oplus \cdots \oplus a_s \odot v^s \mid  a_1,\ldots,a_s \in \R \}.$$
A {\em tropical line segment} between two points $v^1, \, v^2$ is a tropical polytope of a set of two points $\{v^1, \, v^2\} \subset \mathbb R^e \!/\mathbb R {\bf   1}$.
\end{definition}

\begin{definition}[Generalized Hilbert Projective Metric]
\label{eq:tropmetric} 
For any vectors $v:=(v_1, \ldots , v_e), \, w := (w_1, \ldots , w_e) \in \mathbb R^e \!/\mathbb R {\bf 1}$,  the {\em tropical symmetric metric} $d_{\rm tr}$ between $v$ and $w$ is defined as:
\begin{equation*}
{\color{black}d_{\rm tr}(v,w)  := \max_{i \in [e]} \bigl\{ v_i - w_i \bigr\} - \min_{i \in [e]} \bigl\{ v_i - w_i \bigr\}.}
\end{equation*}
$d_{\rm tr}$ is a well-defined metric over the tropical projective torus $\mathbb R^e \!/\mathbb R {\bf 1}$ \cite{LSTY}. 
\end{definition}

\subsection{Space of Ultrametrics}

Suppose we have a map $d: [m] \times [m] \to \RR$ such that
\[
d(i, j) = \begin{cases}
    d(j, i) & \mbox{for all } i, j \in [m]\\
    0 & \mbox{if and only if } i = j
\end{cases}
\]
and 
\[
d(i, j) \leq d(i, k) + d(j, k) \mbox{ for all }i, j, k \in [m].
\]

Then we say that the map $d$ is metric.  
Suppose $d$ is a metric on $[m]$.  Then if $d$ satisfies the following stronger triangle inequalities such that
\begin{eqnarray}
\max\{d(i, j), d(i, k), d(j, k)\} 
\end{eqnarray}
is attained at least twice for any  {\color{black} $i,j,k \in [m]$ then, a metric $d$ is called an {\em ultrametric}. } 

\begin{example}
Suppose $m = 3$.  Let $d$ be a metric on $[m]:=\{1, 2, 3\}$ such that
\[
d(1, 2) = 2, \, d(1, 3) = 2,\, d(2, 3) = 1.
\]
Since the maximum is achieved twice, $d$ is an ultrametric.
\end{example}

A phylogenetic tree is a non-negatively weighted (rooted or unrooted) tree such that their internal nodes do not have labels, their external leaves have labels $[m]$, and each edge in the phylogenetic tree has non-negative weight.  Recall that, in this paper, we consider rooted phylogenetic trees with a leaf label set $[m]$.  
\begin{definition}
Suppose we have a rooted phylogenetic tree $T$ with a leaf label set $[m]$.  If the total branch weight on the unique path from its root to each leaf $i \in [m]$ {\color{black}is  the same} for all $i \in [m]$, then we call $T$ an {\em equidistant tree}.
\end{definition}

{\color{black}The MSC model is the most widely used model for statistical analysis of gene trees. A key assumption of the MSC is that all gene trees are equidistant. Accordingly, we focus on equidistant trees in this paper.}

Note that a phylogenetic tree is a discrete object so in order to conduct a statistical analysis, typically we convert a phylogenetic tree to a vector representation.  Here we convert a phylogenetic tree to a {\em dissimilarity map}.  Dissimilarity maps are maps $d: [m] \times [m] \to \RR$ such that $d(i, i) = 0$ and $d(i, j) = d(j, i)$.  In phylogenetics, we convert a phylogenetic tree to a dissmilarity map by setting $d(i, j)$ equal to the pairwise distance, the total weight on the unique path, between a leaf $i \in [m]$ {and} a leaf $j \in [m]$.  
Then we have the following theorem:

\begin{theorem}[\cite{Buneman}]\label{thm:3pt}
Suppose we have an equidistant tree $T$ with a leaf label set $[m]$ and suppose $d(i, j)$ for all $i, j \in [m]$ is the distance from a leaf $i$ to a leaf $j$.  Then, $d$ is an ultrametric if and only if $T$ is an equidistant tree. 
\end{theorem}

Using Theorem \ref{thm:3pt}, we consider the space of ultrametrics as the space of phylogenetic trees on $[m]$.  Here we define $\Tn$ as the space of ultrametrics with a set of leaf labels $[m]$.  Ardila and Klivans characterized $\Tn$ as the {\em tropical linear space} which is tropically convex.  

\begin{theorem}[\cite{AK,10.1093/bioinformatics/btaa564}]\label{th:tropicalConvex}
Suppose we have a {\em classical} linear subspace $L_m \subset \RR^e$ defined by the linear
equations $x_{ij} - x_{ik} + x_{jk}=0$ for $1\leq i < j <k \leq m$. Let $\mbox{Trop}(L_m)\subseteq \RR^e/\RR {\bf 1}$ be the {\em tropicalization} of the linear space $L_m \subset \RR^e$, that is, we replace the classical addition by the tropical addition $\boxplus$ and we replace the classical multiplication by the tropical multiplication $\odot$ in the equations defining the linear subspace $L_m$, so that all points $(x_{12},x_{13},\ldots, x_{m-1,m})$ in $\Trop(L_m)$ satisfy the condition:
\[
\max_{i,j,k\in [m]}\{v_{ij},v_{ik},v_{jk}\}
\]
{is achieved at least twice}. 
Then the image of $\mathcal U_m$ inside of the tropical projective torus $\RR^e/\RR {\bf 1}$ is equal to $\Trop(L_m)$. 
\end{theorem}

Since $\Tn$ is tropically convex by Theorem \ref{th:tropicalConvex}, Yoshida et al.~in \cite{tropicalKDE} introduced the tropical KDE an analogue of a classical KDE with the tropical symmetric metric $d_{\rm tr}$ defined in {\color{black}Section \ref{sec:tropicalKDE}}.  

\section{Tropical Kernel Density Estimation}\label{sec:tropicalKDE}

Let $\mathcal{S}:= \{T_1, \ldots, T_N \}\subset \Tn$ be a random sample from the space of ultrametrics $\Tn$. In 2024, Yoshida et al. \cite{tropicalKDE} proposed a {\em tropical kernel density estimator (KDE)} with the 
tropical metric over the space of ultrametrics $\Tn$ as an analogue of a classical KDE defined as:
\begin{equation}\label{mastereq}
\hat{f}(T) \propto \frac{1}{N} \sum_{i=1}^N k_{\sigma_i}(T, T_i)
\end{equation}
where $k$ is a non-negative function defined over $\Tn$ such that 
\begin{equation}\label{eq:kernel}
    k_{\sigma_i}(T, T_i) = \exp \left( { - {\left({\frac{d_{\rm tr}(T,T_i)}{\sigma_i}}\right)}}\right) ,
\end{equation}
{\color{black}with a ``bandwidth'' parameter controls the local smoothness of a given kernel function $k(T,T_i)$ at any observation $T_i \in \mathcal{S}$ with respect to $d _{\rm tr}$.} Yoshida et al.~showed by computational experiments that the normalized constant 
\[
C(T_i) = \int_{\Tn} {k_{\sigma_i}(T, T_i)} dT
\]
does not depend on $T_i \in \Tn$ \cite{tropicalKDE}. 
{\color{black}In this paper we make the same assumption.}  Let $\sigma = (\sigma_1, \ldots , \sigma_N)$. 
Then, Yoshida et al.~in \cite{tropicalKDE} considered the estimation
\begin{equation}\label{eq:tropicalKDE1}
    \hat{g}_{\sigma}'(T_j) = \frac{1}{\color{black}(N-1)} \sum_{i \neq j} k_{\sigma_i}(T_j, T_i)
\end{equation}
for $T_j \in \mathcal{S}$.
In this paper{\color{black}, as Yoshida et al.~in \cite{tropicalKDE},} we consider {\color{black}the following:} 
\begin{equation}\label{eq:tropicalKDE}
    \hat{g}_{\sigma}(T_j) = \frac{1}{(N-1)} \sum_{i \neq j}\frac{1}{\sigma_i} k_{\sigma_i}(T_j, T_i)
\end{equation}
for $T_j \in \mathcal{S}$.

\subsection{Likelihood Cross Validation for Kernel Density Estimation}

In \cite{KDE, KDETree, tropicalKDE}, the default set up of this user-defined parameter $\sigma > 0$ is determined by the nearest neighbor of each $T_i \in \mathcal{S}$.  However, this might not be an  {\color{black}optimal selection of the parameter $\sigma $}.  In this paper, we focus on one of the most popular bandwidth {\color{black} selection methods}, the Likelihood Cross Validation (LCV) defined by Habbema et al.~\cite{habbema1974stepwise}:
\begin{eqnarray}\nonumber
    \max_{\sigma > 0} M(\sigma) &=& \max_{\sigma > 0} \frac{1}{N} \sum_{j = 1}^N \ln \left(\hat{g}_{\sigma}(T_j)\right)\\ \label{eq:opt}
    &=& \max_{\sigma > 0} \frac{1}{N} \sum_{j = 1}^N \ln\left(\frac{1}{(N-1)} \sum_{i \not = j} \frac{1}{\sigma_i}\exp\left(\frac{-d_{\rm tr}(T_j, T_i)}{\sigma_i}\right)\right). 
\end{eqnarray}

Now we take a partial derivative on $M(\sigma)$ in terms of $\sigma$ using the {\color{black}product law}:
\[
\frac{\partial M(\sigma)}{\partial \sigma_i} = \frac{\partial M(\sigma)}{\partial \hat{g}_{\sigma}(T_j)}\cdot \frac{\partial \hat{g}_{\sigma}(T_j)}{\partial \sigma_i} ,
\]
for each $i = 1, \ldots, N$ and 
where 
\[
\frac{\partial M(\sigma)}{\partial \hat{g}_{\sigma}(T_j)} = \frac{(N-1) }{{\sum_{i \not = j} \frac{1}{\sigma_i}\exp\left(\frac{-d_{\rm tr}(T_j, T_i)}{\sigma_i}\right)}}
\]
and 
\[
\frac{\partial \hat{g}_{\sigma}(T_j)}{\partial \sigma_i} = \frac{1}{(N-1)} \frac{\left(\sigma_i - d_{\rm tr}(T_j, T_i)\right)\exp\left(\frac{-d_{\rm tr}(T_j, T_i)}{\sigma_i}\right)}{\sigma_i^2}. 
\]
Therefore we have
\[
\frac{\partial M(\sigma)}{\partial \sigma_i} = \frac{\partial M(\sigma)}{\partial \hat{g}_{\sigma}(T_j)}\cdot \frac{\partial \hat{g}_{\sigma}(T_j)}{\partial \sigma_i} = \frac{\left(\sigma_i - d_{\rm tr}(T_j, T_i)\right)}{\sigma_i^2}\cdot \frac{\exp\left(\frac{-d_{\rm tr}(T_j, T_i)}{\sigma_i}\right)}{{\sum_{i \not = j} \frac{1}{\sigma_i}\exp\left(\frac{-d_{\rm tr}(T_j, T_i)}{\sigma_i}\right)}}
\]
for each $i = 1, \ldots, N$.  Thus, when 
\[
\sigma_i = d_{\rm tr}(T_j, T_i)
\]
for each $i = 1, \ldots, N$,  we have
\[
\frac{\partial M(\sigma)}{\partial \sigma_i} = 0.
\]
Therefore,
the equation in \eqref{eq:opt} {\color{black}achieves optimal value}.  


\section{Computational Experiments}

In this section we conduct two computational experiments with {\color{black} modified tropical KDE:} {\color{black}(1) computational experiments with simulated data generated from the MSC model with a species tree, and (2) computational experiments with empirical data from the Apicomplexa genome dataset \cite{kuo}.}
We apply the same experimental design from Yoshida et al.~in \cite{tropicalKDE}.

\subsection{Multi-Species Coalescent Model}
For computational experiments with simulated data, we use a software {\tt Mesquite} \cite{mesquite} which generates gene trees with a given species tree under the MSC model.  Under the MSC model with a fixed species tree, there are two parameters: the effective population size $N_e$ and the species depth (SD).  Under the MSC model, the effective population size $N_e$ is the population size for each species and the SD is the number of generations from the most recent common ancestor. Under this computational experiment, we fix $N_e = 100,000$ and we vary 
$$R = \frac{SD}{N_e}.$$
Under this experiment, we fix the number of leaves $m = 10$.  See Algorithm 1 in \cite{tropicalKDE} for more details on a way to generate simulated dataset.
In these simulated
experiments, we vary the ratio $R = 0.25, 0.5, \, 
1, \, 2, \, 5, \, 10$.
\begin{remark}
\normalsize
    When we have small $R$, gene trees generated under a MSC model with a given species tree have wider variations, so that they tend to behave like random trees.  Thus, it becomes harder to distinguish between two distributions of gene trees with two different MSC model with different species trees as $R$ becomes smaller \cite{rannala:hal-02535622}. {\color{brown}\textbf{Comment: font fixed}}
\end{remark}
 
Let $T_1$ and $T_2$ be species trees generated by the Yule model with a leaf label set $[m]$.  Let $\mathcal{S}_1$ be a sample with sample of gene trees generated under the MSC model with the species tree $T_1$ with a size $N$ and  $\mathcal{S}_2$ be a sample with sample of gene trees generated under the MSC model with the species tree $T_2$ with a size $N$.  Let $n_1 > 1$ be the number of non-outlier gene trees and $n_2 > 1$ be the number of outlier gene trees.  Let $M$ be a KDE for gene trees.  The following algorithm is from Yoshida et al.~in \cite{tropicalKDE}.

\begin{algorithm}
\caption{Experiments on a Sample Generating from MSC (Algorithm 1 in \cite{tropicalKDE})}\label{alg:sim1}
\begin{algorithmic}
\State {\bf Input:} $n_1 > 1$ many non-outlier gene trees $T_1, \ldots , T_{n_1}$; and $n_2 \geq 1$ many outlier gene trees $T'_1, \ldots , T'_{n_2}$.  Density Estimator $M$.
\State {\bf Output:} Estimated probabilities for $n_1$ many non-outlier gene trees and $n_2$ many outlier gene trees.
\For{$j= 1, \ldots , n_2$,}
\For{$i= 1, \ldots , n_2$,}
\State Compute estimated probability $\hat{f}(T_i)$ of $T_i$ via $M$ with a sample of gene trees $\{T_1, \ldots , T_{i-1}, T_{i+1}, \ldots , T_{n_1}, T'_j\}$. 
\State Compute estimated probability $\hat{f}(T'_j)$ of $T'_j$ via $M$ with a sample of gene trees $\{T_1, \ldots , T_{n_1}\}$. 
\EndFor 
\EndFor \\
\Return $\hat{f}(T_1), \ldots , \hat{f}(T_{n_1})$ and $\hat{f}(T'_1), \ldots , \hat{f}(T'_{n_2})$.
\end{algorithmic}
\end{algorithm}

\begin{figure*}
    \centering
    \includegraphics[width=0.8\textwidth]{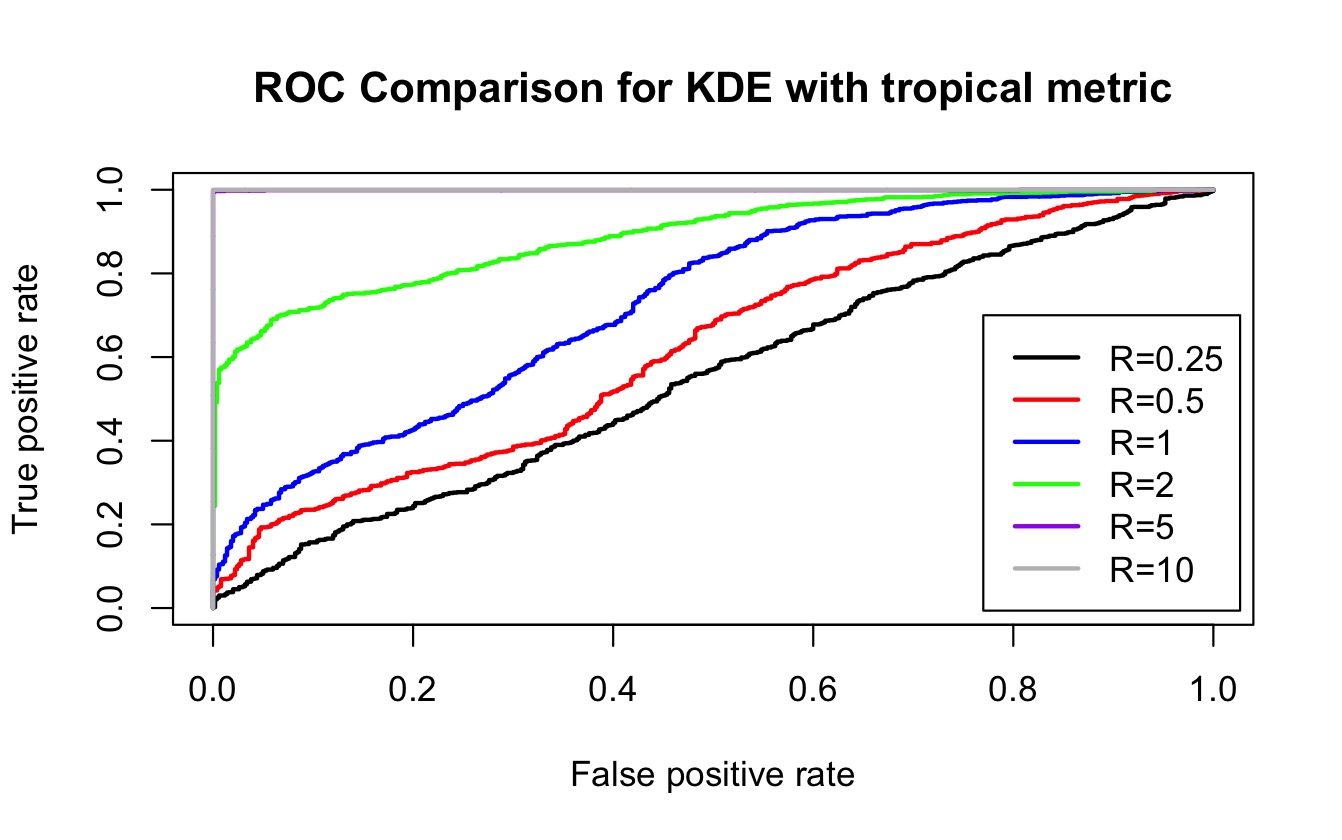}
    \caption{ROC curves for the KDE with the tropical metric. }
    \label{fig:KDE_ROC}
\end{figure*}

Similar to Yoshida et al.~in \cite{tropicalKDE},  
we apply Algorithm \ref{alg:sim1} with $n_2 = 500$ and $n_1 = 1000$.  More specifically, for each $R$, we take all $1000$ trees from $\mathcal{S}_1$ and we take one tree from $\mathcal{S}_2$.  
{\color{black}Then we estimate the probability distribution of gene trees using the tropical density estimator described in Equation~\eqref{eq:tropicalKDE}, and we also estimate it via {\tt KDETrees} \cite{tropicalKDE}.} We iterate this process $500$ times.  Therefore, we have estimated probabilities for $1000$ trees in $\mathcal{S}_1$ and for $500$ trees in $\mathcal{S}_2$. 
The ROC curves for the output from the simulation are shown in Fig.~\ref{fig:KDE_ROC} and AUCs are shown in Table \ref{tab:AUC}. Note that these results are {\color{black}slightly better than} results in \cite{tropicalKDE}.  {\color{black}In addition, the computational time of our estimator is much faster than it is in \cite{tropicalKDE}.  More specifically the computational time of our estimator takes $13.11$ minutes per each $R$ in this computational experiment while the computational time of the estimator in \cite{tropicalKDE} takes $28.27$ minutes per each $R$.}
\begin{table}{Area Under the Curves (AUCs)}
    \centering
    \begin{tabular}{|c|cccccc|}\hline
    $R$ & 0.25 & 0.5 & 1 & 2 & 5 & 10\\\hline
    {\bf LCV} & {\bf $0.55$}& {\bf $0.62$}& {\bf $0.73$}& {\bf $0.89$}& {\bf $1.00$}&{\bf   $1$}\\
         NN &$0.54$& $0.61$& $0.71$& $0.88$& $1.00$&  $1$\\\hline
    \end{tabular}
    \caption{Area Under the Curves (AUCs) for the tropical KDE with the optimal LCV and with the nearest neighbors (NN) proposed by Yoshida et al.~\cite{tropicalKDE}. }
    \label{tab:AUC}
\end{table}

\begin{table}{Likelihood Cross Validation (LCV) Values}
    \centering
    \begin{tabular}{|c|ccccc|}\hline
    $R$ & 0.5 & 1 & 2 & 5 & 10 \\\hline
    {\bf LCV} & {\bf $-1.38$} & {\bf $-1.28$} & {\bf $-0.98$} & {\bf $-0.39$} & {\bf $0.21$} \\
    \hline
    \end{tabular}
    \caption{Likelihood Cross Validation (LCV) values for the tropical kernel density estimator under various levels of gene tree discordance ($R$).  Note that in our computation estimated probability  $\hat{g}_{\sigma}$ is not normalized.}
    \label{tab:LCV}
\end{table}

\subsection{Apicomplexa}
In this section we apply our tropical KDE to the Apicomplexa dataset which consists of  268  orthologous sequences  with  eight  species  of  protozoa  generated by Kuo et al.~\cite{kuo}.  The dataset contains eight species in each alignment in the set: {\it Babesia  bovis} (Bb), {\it Cryptosporidium
  parvum} (Cp), {\it Eimeria tenella} (Et), {\it
  Plasmodium falciparum} (Pf), {\it Plasmodium  vivax} (Pv),
{\it Theileria  annulata} (Ta),  and {\it Toxoplasma
  gondii} (Tg).  Among them the outgroup is a free-living ciliate, {\it
  Tetrahymena  thermophila} (Tt).

We compute gene trees in the $0.05$ lower tail of the estimated distribution of gene trees using our kernel density estimator. Compared with the result from Yoshida et al.~\cite{tropicalKDE}, our estimator did not include two trees with the IDs: 630 and 503.  The gene trees which both of our KDE and the KDE from \cite{tropicalKDE} identify as in the $0.05$ lower tail of the estimated distribution of gene trees are trees with their IDs 691, 566, 650, 730, 615, 712, 625, 755, 708, 497, 690 (ordered by the smallest probabilities to the largest) whose detailed information is listed in Table \ref{tab:outlyinggenes}.
\begin{table}{Apicomplexa gene sets identified as outliers by KDE with the tropical metric}
    \centering
    \begin{tabular}{|c|c|p{8cm}|}\hline
        \# & Gene ID & Function \\\hline
         691 & PFA0310c & calcium-transporting ATPase\\
         566 & PF13\_0257 & glutamate--tRNA ligase\\
         650 & PF11\_0358 & DNA-directed RNA polymerase, beta subunit, putative\\
         730& PFL0930w & clathrin heavy chain, putative\\
         615 & PF13\_0063 & 26S proteasome regulatory subunit 7, putative\\
         712 & MAL13P1.274 & serine/threonine protein phosphatase pfPp5\\
         625 & PFD1090c & clathrin assembly protein, putative\\
         755& PF10\_0148 & hypothetical protein\\
         708 & PFC0140c & N-ethylmaleimide-sensitive fusion protein, putative\\
         497 & PF13\_0228 & 40S ribosomal subunit protein S6, putative\\
         690 & MAL8P1.134 & hypothetical protein, conserved\\\hline
    \end{tabular}
    \caption{Apicomplexa gene sets identified as outliers by KDE with the tropical metric.  All annotations except 728 are putative. Based on the gene set designations in \cite{kuo}. Gene set represented by GeneID for {\it P.falciparum}.} 
    \label{tab:outlyinggenes}
\end{table}

\section{Discussion}

In this paper we propose LCV to estimate the optimal bandwidth for tropical KDEs for estimating the distribution of gene trees over the space of ultrametrics.  It is interesting to note that from the computational results, the NN for estimating the optimal bandwidth proposed by Yoshida et al.~\cite{tropicalKDE} seems to achieve almost the optimal in terms of LCV.  

When we have the bandwidth 
\[
\sigma_i = d_{\rm tr}(T, T_i),
\]
one notices that we have
\[
k_{\sigma_i}(T, T_i) = \exp \left( { - {\left({\frac{d_{\rm tr}(T,T_i)}{\sigma_i}}\right)}}\right) = \exp(-1).
\]
This simplifies the kernel function and computation.  However, this is only the optimal bandwidth in terms of the LCV.  
{\color{black}Another interesting approach is the Least Squares Cross Validation (LSCV) proposed by Rudemo \cite{LSCV}.}

\bmhead{Acknowledgements}
RY is partially supported from NSF DMS 2409819.  Also RY participated and is supported by the 2023 American Women in Mathematics (AWM) Symposium at the Clark Atlanta University on September 30th to October 2nd 2023.

\section*{Declarations}

Some journals require declarations to be submitted in a standardised format. Please check the Instructions for Authors of the journal to which you are submitting to see if you need to complete this section. If yes, your manuscript must contain the following sections under the heading `Declarations':

\begin{itemize}
\item Funding: RY is partially supported from NSF DMS 2409819.
\item Conflict of interest/Competing interests: No conflicts of interests.
\item Ethics approval and consent to participate: Not applicable.
\item Consent for publication:  The authors agreed to consent for publication.  
\item Data availability: \url{https://polytopes.net/tropical_KDE_new.zip}
\item Materials availability: \url{https://polytopes.net/tropical_KDE_new.zip}
\item Code availability: \url{https://polytopes.net/tropical_KDE_new.zip}
\item Author contribution:  RY contributes the problem setting and theoretical results.  ZW contributes computational results.  
\end{itemize}

\bibliography{refs}

\end{document}